\documentclass[preprint]{aastex631}
\shorttitle{Mapping Turbulence in a Solar Flare}
\shortauthors{Stores et al.}
\graphicspath{{./}{figures/}}

\begin{document}

\title{The Spatial and Temporal Variations of Turbulence in a Solar Flare}

\author[0000-0002-6060-8048]{Morgan Stores}
\affiliation{Department of Mathematics, Physics \& Electrical Engineering, Northumbria University,  \\
Newcastle upon Tyne, UK\\
NE1 8ST}

\author[0000-0001-6583-1989]{Natasha L. S. Jeffrey}
\affiliation{Department of Mathematics, Physics \& Electrical Engineering, Northumbria University,  \\
Newcastle upon Tyne, UK\\
NE1 8ST}

\author[0000-0002-8078-0902]{Eduard P. Kontar}
\affiliation{School of Physics \& Astronomy, University of Glasgow, \\
Glasgow, UK \\
G12 8QQ}



\begin{abstract}

Magnetohydrodynamic (MHD) plasma turbulence is believed to play a vital role in the production of energetic electrons during solar flares and the non-thermal broadening of spectral lines is a key sign of this turbulence. Here, we determine how flare turbulence evolves in time and space using spectral profiles of Fe \textsc{xxiv}, Fe \textsc{xxiii} and Fe \textsc{xvi}, observed by \emph{Hinode}/EIS. Maps of non-thermal velocity are created for times covering the X-ray rise, peak, and decay. For the first time, the creation of kinetic energy density maps reveal where energy is available for energization, suggesting that similar levels of energy may be available to heat and/or accelerate electrons in large regions of the flare. We find that turbulence is distributed throughout the entire flare; often greatest in the coronal loop tops, and decaying at different rates at different locations. For hotter ions (Fe \textsc{xxiv} and Fe \textsc{xxiii}), the non-thermal velocity decreases as the flare evolves and during/after the X-ray peak shows a clear spatial variation decreasing linearly from the loop apex towards the ribbon. For the cooler ion (Fe \textsc{xvi}), the non-thermal velocity remains relativity constant throughout the flare, but steeply increases in one region corresponding to the southern ribbon, peaking just prior to the peak in hard X-rays before declining. The results suggest turbulence has a more complex temporal and spatial structure than previously assumed, while newly introduced turbulent kinetic energy maps show the availability of the energy and identify important spatial inhomogeneities in the macroscopic plasma motions leading to turbulence.
\end{abstract}

\keywords{Sun: flares – Sun: chromosphere – Sun: corona – Sun: UV radiation - turbulence - techniques: spectroscopic}

\section{Introduction} \label{sec:intro}

During a solar flare, energy stored in twisted and stressed coronal magnetic fields is rapidly released by magnetic reconnection \citep[e.g.,][]{1957JGR....62..509P, 1958IAUS....6..123S,2000mare.book.....P}, resulting in low lying flare loops. The magnetic energy released by reconnection is partitioned into thermal and non-thermal particle energies \citep[e.g.,][]{2012ApJ...759...71E,warmuth2016constraints,2017ApJ...836...17A}, as well as kinetic energy in the form of plasma motions and turbulence in the coronal loop tops of flares \citep{1994ApJS...90..623M,2017PhRvL.118o5101K}.
 
Magnetohydrodynamic (MHD) plasma turbulence (stochastic motions of the magnetized plasma), a product of reconnection, can drastically affect the dynamics of the reconnection process itself and transfer energy facilitating particle energization and heating over 10 MK \citep[e.g.,][]{larosa1993mechanism,petrosian2012stochastic,2017PhRvL.118o5101K} via the cascade of energy to smaller and smaller scales \citep{2017SSRv..212.1107K} and eventually dissipation at the particle level \citep[e.g.,][]{miller1997critical,krucker2008hard,petrosian2012stochastic}. This rate of particle energization is determined by both the rate of energy release at large scales and the rate of energy transfer to smaller scales. Within these regions of turbulence, charged particles can also become trapped by turbulent scattering off magnetic fluctuations, particularly high-energy MeV electrons \citep{musset2018diffusive}. Thus, the presence of turbulence in flares is intimately linked with both electron acceleration and transport.
The properties of the solar flare plasma can be determined from observing and analyzing the properties of optically thin spectral lines in the solar atmosphere \citep[e.g., see reviews by][]{2015SoPh..290.3399M,2018LRSP...15....5D} with instruments such as the \emph{Hinode} \citep{2007SoPh..243....3K} EUV Imaging Spectrometer (EIS) \citep{2007SoPh..243...19C} providing spatially resolved information on ion line emissions, plasma temperatures, mass flows, ion abundances, and electron densities. Spectral line width or broadening is produced by many small shifts in wavelength caused by both microscopic ion motions and macroscopic plasma motions of the emitting ion. During a flare, spectral lines often show line widths in excess of what is expected from random thermal motions alone (thermal broadening or velocity). The excess broadening (non-thermal broadening or velocity) is believed to be the result of superposed Doppler-shifted emission of (1) turbulent fluid motions and/or (2) the superposition of flows from different loops. Recently \citet{2019ApJ...879L..17P}, using hydrodynamic simulations of multi-thread flare loops with the 1D RADYN code \citep{1992ApJ...397L..59C,2015ApJ...809..104A}, suggested that non-thermal broadening cannot be easily reproduced by such flows, with turbulent motions most likely to be the cause. Alternative mechanisms may cause line broadening, such as opacity or pressure broadening. However as the solar flare plasma is optically thin \citep[e.g.,][]{levens2018diagnostics}, these should be negligible \citep[e.g.,][]{2011ApJ...740...70M}. Spectral line broadening is usually extracted by fitting a Gaussian profile to the line, as such a line shape is expected from a thermal ion velocity distribution and indeed from any plasma motions undergoing Brownian motion. However, it has also been suggested that the line shape may provide a diagnostic of the underlying turbulent plasma motions \citep[e.g., ][]{jeffrey2016first,2017ApJ...836...35J,2018ApJ...864...63P} where non-Brownian plasma motions or the spectrum of plasma/magnetic fluctuations has a non-Gaussian or Lorentzian form \citep[e.g., ][]{2014ApJ...780..176K} then different spectral line shapes should indeed be expected. 
 
Non-thermal broadening has been extensively studied in solar flares and hence serves as a key signature of turbulence. Early studies using soft X-rays (SXR) were spatially unresolved \citep[e.g.,][]{1980ApJ...239..725D,1982SoPh...78..107A,1995ApJ...438..480A}, but they showed: (1) non-thermal motions preceding the convective motions caused by chromospheric evaporation suggesting that non-thermal broadening is due to the primary energy release and (2) a strong temperature dependence where regions of high temperature and hence lines formed at the highest temperatures show the greatest non-thermal broadening. More recently, spatial changes in non-thermal broadening have also been examined \citep[e.g.,][]{2014ApJ...788...26D,2018ApJ...854..122W} using \emph{Hinode} EIS and also with the \emph{Interface Region Imaging Spectrograph (IRIS)} \citep{2014SoPh..289.2733D}. Spatially resolved observations often show that non-thermal broadening increases with both temperature and height in the coronal loop source. Interestingly, although turbulence is often discussed as a product of reconnection in the corona, the presence of non-thermal broadening during the flare is not a phenomenon only localized high in the coronal loops. For example, \citet{2011ApJ...740...70M} looked at both bulk Doppler flows and non-thermal broadening in cooler lines formed in the low corona close to flare X-ray footpoints. \citet{2018SciA....4.2794J} studied Si \textsc{iv} formed in the lower transition region during a flare. The high time cadence observations using \emph{IRIS} showed that non-thermal broadening increased a full 10 seconds before heating in the region, strongly suggesting turbulence played a role in the heating. Moreover, the temporal evolution of both bulk velocity and non-thermal velocity could be reproduced by plasma fluctuations caused by the interaction of different wave forms in the region.
 
In this study, we perform a thorough investigation of the spatial and temporal distribution of non-thermal spectral line broadening (turbulence) in one flare. We map how the non-thermal broadening and kinetic energy of plasma motions change throughout the entire flare in order to determine how the distribution and magnitude of turbulence changes with temperature, time and space. The non-thermal broadening and turbulent kinetic energy from various ions is mapped using EUV spectral lines from \emph{Hinode} EIS formed between $\approx$6.3 to 15.8 MK. High resolution EUV spatial images from the Atmospheric Imaging Assembly (AIA) \citep{2012SoPh..275...17L} aboard the \emph{Solar Dynamic Observatory (SDO)} are used in conjunction with EIS data to map the non-thermal broadening to flare features, such as the coronal loop top source and the ribbon features, and to different loops produced as reconnection progresses. An overview of the data is provided in Sect. \ref{overview}. Sect \ref{results} presents the results and the findings of the study are summarized in Sect. \ref{sec:discussion}.
 
\section{Observational Overview}
\label{overview}

\begin{figure}
\centering
\includegraphics{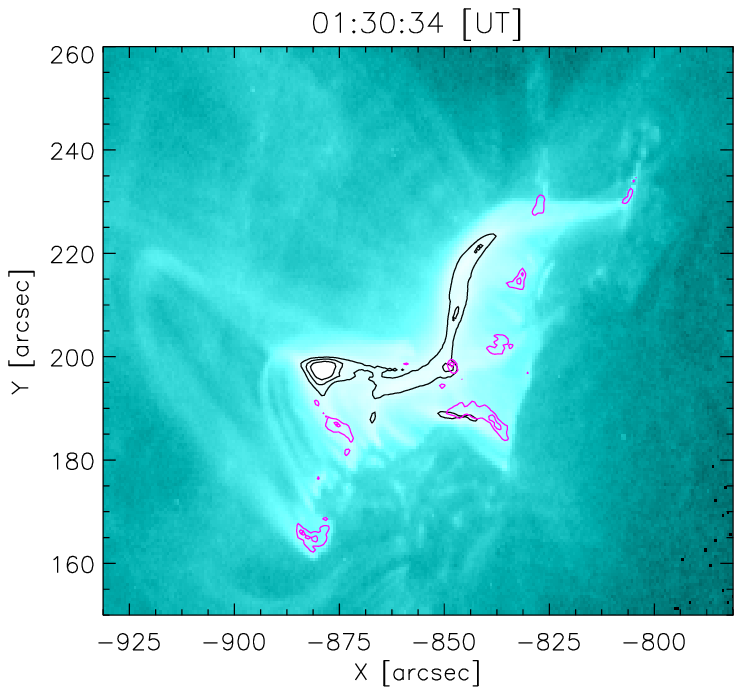}
\includegraphics{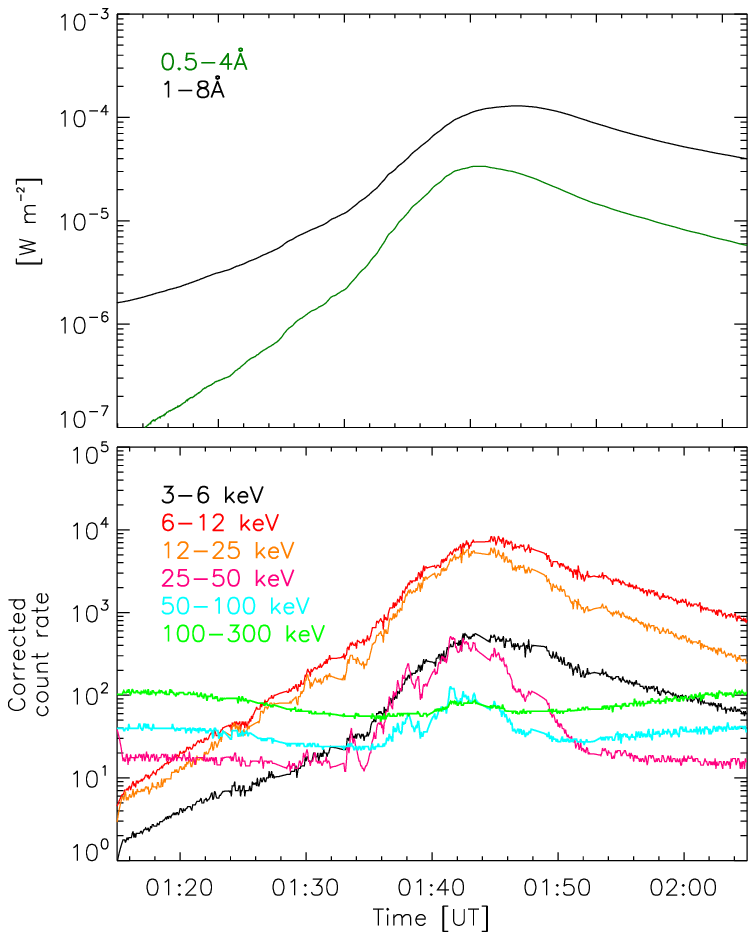}
\caption{AIA 131 \AA~ image (\textit{top}) taken at 01:30:34 UT of the X1.2 flare SOL2013-05-15T01:45. Black contours show AIA channel 131 \AA~intensity levels at 50$\%$, 70$\%$, and  90$\%$ of the maximum 131 \AA~image value (units of DN s$^{-1}$ px$^{-1}$). Pink contours show AIA channel 304 \AA~intensity levels at 40$\%$ and 80$\%$ of the maximum 304 \AA~image value. \emph{GOES} and \emph{RHESSI} SXR and HXR (\textit{middle and bottom}) flare light curves, peaking at approximately 01:45:00 UT and 01:42:00 UT respectively. These observations coincide with the time of EIS observations that range from 01:25:14 UT to 01:51:58 UT.}
\label{flare}
\end{figure}

This paper studies the X1.2 solar flare, SOL2013-05-15T01:45 shown in Figure \ref{flare}. The \emph{Geostationary Operational Environmental Satellite (GOES)} \citep{1991SoPh..133..371K} light curves in Figure \ref{flare} show the SXR emission, corresponding to the heated plasma, peaking at approximately 01:45 UT. Light curves from the 
\emph{(Reuven) Ramaty High Energy Solar Spectroscopic Imager (RHESSI)} \citep{2002SoPh..210....3L}, see Figure \ref{flare}, show the hard X-ray (HXR) emission, usually corresponding to accelerated electrons, rising and peaking at approximately 01:34:00 UT and 01:42:00 UT respectively. 
To study different flare features, AIA images at several wavelengths were examined, including 131 \AA~and 304 \AA. Figure \ref{flare} shows an AIA 131 \AA~ image of the flare at 01:30:34 UT, with 131 \AA~contours (black) at levels of 50$\%$, 70$\%$, and 90$\%$ of the maximum 131 \AA~image value, and 304 \AA~contours (pink) at levels of 40$\%$ and 80$\%$, of the maximum 304 \AA~image value (units of DN s$^{-1}$ px$^{-1}$). Cooler channels such as 304 \AA~highlight the ribbon features of the flare, whilst the hotter channels such as 131 \AA~highlight the coronal source and hot flaring loops. 

\subsection{Hinode EIS Spectral Data}
Prior to this solar flare, there were multiple large flares in the active region. This study utilizes \emph{Hinode} EIS spectral line data between the times of 01:30:34 UT and 01:51:58 UT, covering the rise, the peak, and part of the decay phase of the flare, as shown in Figure \ref{flare}. During this time EIS operated using the $2\arcsec$ slit in fast rastering mode. This results in a reduced spatial resolution in the $x$ direction, $\sim 5.99\arcsec$ and a temporal resolution of approximately 9~s. The EIS field of view observing the flare region is $(30 \times 5.99\arcsec$) $\sim 179\arcsec \times 152\arcsec$. 

During the studied time frame, five EIS observations are available and each set of data contains 10 spectral windows. Here, the stated time of an observation is at the beginning of the raster time. SolarSoftWare (SSW) routine eis\_prep.pro converted the raw EIS data to level 1 data. This routine removes saturated data, detector artifacts, cosmic ray hits, and missing data packets. 
In order to study the line profiles, strong spectral lines are required that have no blending with other lines. The spectral lines investigated are given in Table \ref{table1} where the data was obtained from the CHIANTI database \citep{1997A&AS..125..149D,2019ApJS..241...22D}. These lines were initially chosen for their high intensity and high formation temperatures, forming in different regions of the flaring corona.

 \begin{table}
    \centering
    \caption{\emph{Hinode}/EIS spectral lines studied.}
    \begin{tabular}{c c c c c }
    \hline
    \hline    
         & Wavelength & Temperature  & Thermal velocity \\ 
         Ion & (\AA) & log$_{10}T $(K)  & (km$\,$s$^{-1}$) \\ 
        \hline 
        Fe \textsc{xxiv} & 192.0285 & 7.2  & 70.0 \\ 
        Fe \textsc{xxiv} & 255.1136 & 7.2  & 70.0 \\
        Fe \textsc{xvi} & 262.9760  & 6.8  & 44.1\\         
        Fe \textsc{xxiii} & 263.7657 & 7.2  & 70.0\\
       \hline
    \end{tabular}
    \label{table1}
\end{table}

Due to the high cadence of AIA data, there is image data available that correlates to the time of the EIS data observations within 10 seconds. To align EIS with the AIA images, EIS Fe \textsc{xxiv} (192.0285 \AA) images were aligned to AIA 193 \AA\footnote{Fe \textsc{xxiv} (192.0285 \AA) accounts for around 80$\%$ of the total intensity of AIA 193 \AA~\citep{2017PhRvL.118o5101K}}. The routine eis\_aia\_offsets.pro was also used to check the alignment between the two data types. This routine has an uncertainty of $\sim 5 \arcsec$  which compared to the EIS field of view is small. Thus, the maximum alignment uncertainty is $\sim 5 \arcsec$.

\subsection{EIS Spectral Line Fitting and Analysis}
This study only examines optically thin spectral lines. Thus, properties such as line shape and broadening can be related to the underlying ion velocity distribution that is assumed to be an isothermal Maxwellian ion distribution\footnote{The study of \citet{jeffrey2016first} determined non-Gaussian line shapes in certain regions of this flare, but the isothermal Maxwellian assumption is appropriate for the study outlined here}. Hence, all studied lines with intensity $I(\lambda)$ are fitted with the following Gaussian function,

\begin{equation}
    I(\lambda) = I_{B} + I_{0}\;\text{exp}\bigg ( -\frac{(\lambda - \lambda_{0})^2}{2\Delta\lambda^2} \bigg ) \, , 
    \label{gauss}
\end{equation}

where $\lambda$ is the wavelength (\AA), $I_{B}$ is the background intensity (erg$\,$cm$^{-2} \,$s$^{-1}\,$sr$^{-1}$ \,\AA$^{-1}$), $I_{0}$ is the peak intensity (erg$\,$cm$^{-2} \,$s$^{-1}\,$sr$^{-1}$\,\AA$^{-1}$), $\lambda_{0}$ is the measured centroid position (\AA), and $\Delta\lambda$ is the line broadening (\AA).

\begin{figure}
    \centering
    \includegraphics[scale = 0.85]{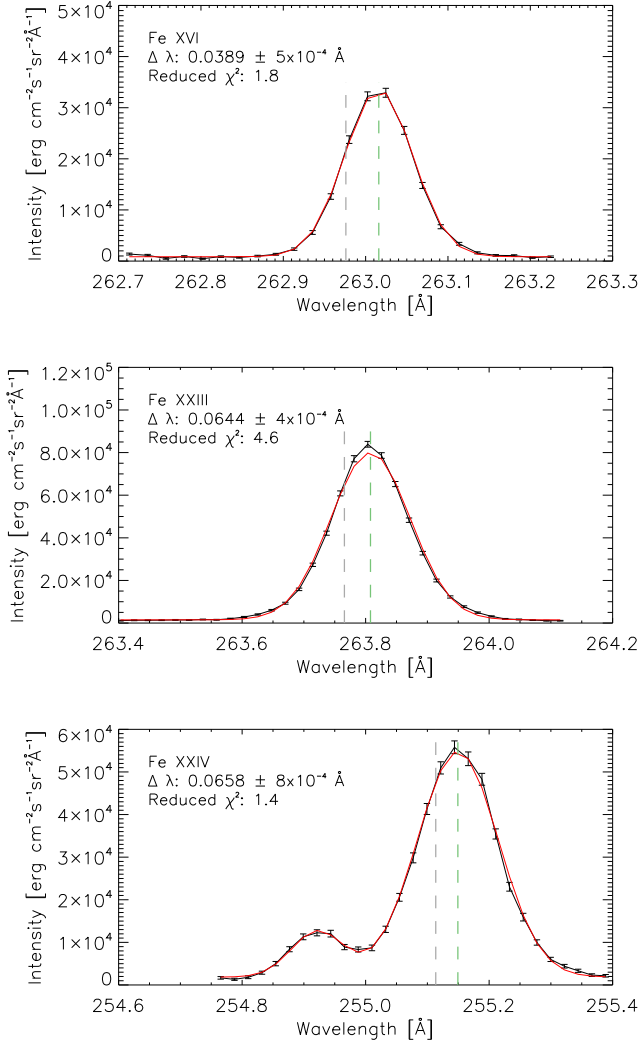}
    \caption{Example Gaussian fits for spectral lines included in the study. \textit{Top to bottom} lines and fits for Fe \textsc{xvi}, Fe \textsc{xxiii}, and Fe \textsc{xxiv} respectively. The red line shows the Gaussian fit while black points show the data. The rest (laboratory) wavelength of the ion is indicated by the gray dashed line in each panel while the observed centroid is shown by the green dashed line and shifted by $0.0400$ \AA, $0.0422$ \AA~and $0.0357$ \AA~for Fe \textsc{xvi}, Fe \textsc{xxiii}, and Fe \textsc{xxiv} respectively. The values of the total line broadening $\pm$ uncertainty ($\Delta\lambda$  [\AA]) and reduced $\chi^2$ value are stated in the legend of each panel. Some Fe \textsc{xxiv} lines required a double Gaussian fit to remove blends (bottom panel).}
    \label{gaussian}
\end{figure}

As shown in Figure \ref{gaussian}, many of the line centroid positions are red-shifted compared to their laboratory rest position, by $\sim0.04$~\AA~respectively. This indicates the presence of bulk motions in the plasma moving away from the observer. In \citet{2017ApJ...836...35J}, a similar observation in a close to disk center flare (SOL2014-03-29T17:44) suggested the presence of plasma down flows alongside spectral line broadening and line shape changes. Indeed, turbulent flows are expected in the flaring corona, due to high Reynolds numbers \citep[e.g.,][]{2014masu.book.....P}.

The spectral lines discussed here are formed over different temperatures: the coolest ion studied is Fe \textsc{xvi} (262.9760 \AA) which according to CHIANTI forms at $\sim6.3$ MK, but the contribution function for Fe \textsc{xvi} peaks at $\sim2-4$~MK \citep[e.g.,][]{jeffrey2016first}; Fe \textsc{xxiii} (263.7657 \AA) which forms at $\sim15.8$ MK; and Fe \textsc{xxiv} (192.0285 \AA~and 255.1136 \AA) also forming at $\sim 15.8$ MK but with a larger contribution function at $\log_{10}T>7.2$ compared to  Fe \textsc{xxiii} \citep[e.g.,][]{2013ApJ...767...83G}.

Analysis of the EIS spectral lines and corresponding Gaussian fits ensured only suitable lines remained in the study. Spectral lines were considered unsuitable and removed from the study if they were very saturated, noisy (defined as the ratio of the intensity uncertainty to intensity), did not have an accurate Gaussian fit, or contained other line components.

Often, a small line component lying in the blue wing of Fe \textsc{xxiii} is observed. There is no documented line at this location in the CHIANTI database and the line could be due to bulk motions along the line of sight. Fe \textsc{xxiii} lines containing the shifted component were removed from the study. Similar components were also seen in the blue wing of Fe \textsc{xvi} (262.9760 \AA), these lines were also removed from the study.

Fe \textsc{xxiv} was originally observed at 192.0285 \AA. Due to high saturation at this wavelength during the flare, this line was discarded. Instead, the Fe \textsc{xxiv} line found at 255.1163 \AA~was studied. The spectral line Fe \textsc{xvii} (254.8853 \AA) sits in the blue wing of Fe \textsc{xxiv} (255.1136 \AA). Due to the spectral resolution of the instrument, 47 m\AA~full width half maximum (FWHM) at 185 \AA, EIS is not always able to resolve blended spectral lines. However, it is possible to separate the emission from each blended component by fitting multiple Gaussian components. A simple two Gaussian fit is successful in eliminating the influence of Fe \textsc{xvii} in the wing.

Solar flare spectral lines have a high intensity compared to background spectral lines. Thus, noise was reduced by applying a lower intensity limit, ensuring that the pixels studied were related to the flare, not background activity. To further reduce the amount of noise, pixels were binned in the $y$-direction.

In order to see how the binning size affected the line broadening, pixels were binned in grouping sizes 2$\arcsec$-5$\arcsec$. Comparing the different binning sizes, we determined no significant change in line broadening with changing binning size. Thus, a binning size of 4$\arcsec$ was found to be suitable; the resulting spectral lines were of high intensity but contained enough spatial information to compare with the flare features.

To determine the accuracy of the Gaussian forward fitting to the spectral line data, the goodness of fit was measured using a standard reduced chi-squared statistic $\chi^2$. 
Reduced $\chi^2$ maps were created for each ion at each observation time and any Gaussian fits with a reduced $\chi^2$ value greater than 5 were extracted and examined carefully by eye.
If the spectral line or Gaussian fit was deemed unsuitable the pixel was removed from the study.


\section{Results}
\label{results}
\subsection{Maps of Non-Thermal Broadening}
To determine the non-thermal broadening from the total line broadening $\Delta\lambda$, firstly, the FWHM (\AA) of each spectral line was calculated using 
\begin{equation}
    \text{FWHM} = 2\sqrt{2\text{ln}2} \,  \Delta\lambda \, .
\end{equation}
A non-thermal velocity $v_{\rm nth}$ (km$\,$s$^{-1}$), associated with the turbulent plasma motions, can be extracted from the line broadening by re-arranging the following equation \citep[e.g.,][]{2016ApJ...820...63B}, that relates the FWHM with the individual thermal, non-thermal and instrumental broadening components,
\begin{equation}
    \text{FWHM} = \sqrt{4\text{ln}2\left ( \frac{\lambda_0}{c}\right )^2\left( \frac{2k_BT_i}{m} + v_{\rm nth}^2\right ) + \text{FWHM}_{\text{I}}^2 } \, ,
    \label{Eq:non-thermal}
\end{equation}
where the thermal velocity is given by $\sqrt{2k_BT_i/m}$ (Table \ref{table1}), and $c$ is the speed of light, $k_B$ is the Boltzmann constant, $T_i$ is the ion temperature (K) and $m$ is the mass of the ion (g). FWHM$_{\text{I}}$ is the instrumental FWHM (\AA) that was determined using SolarSoftWare (SSW) routine eis$\_$slit$\_$width.pro. 

\begin{figure*}
    \centering
    \includegraphics[scale = 0.84]{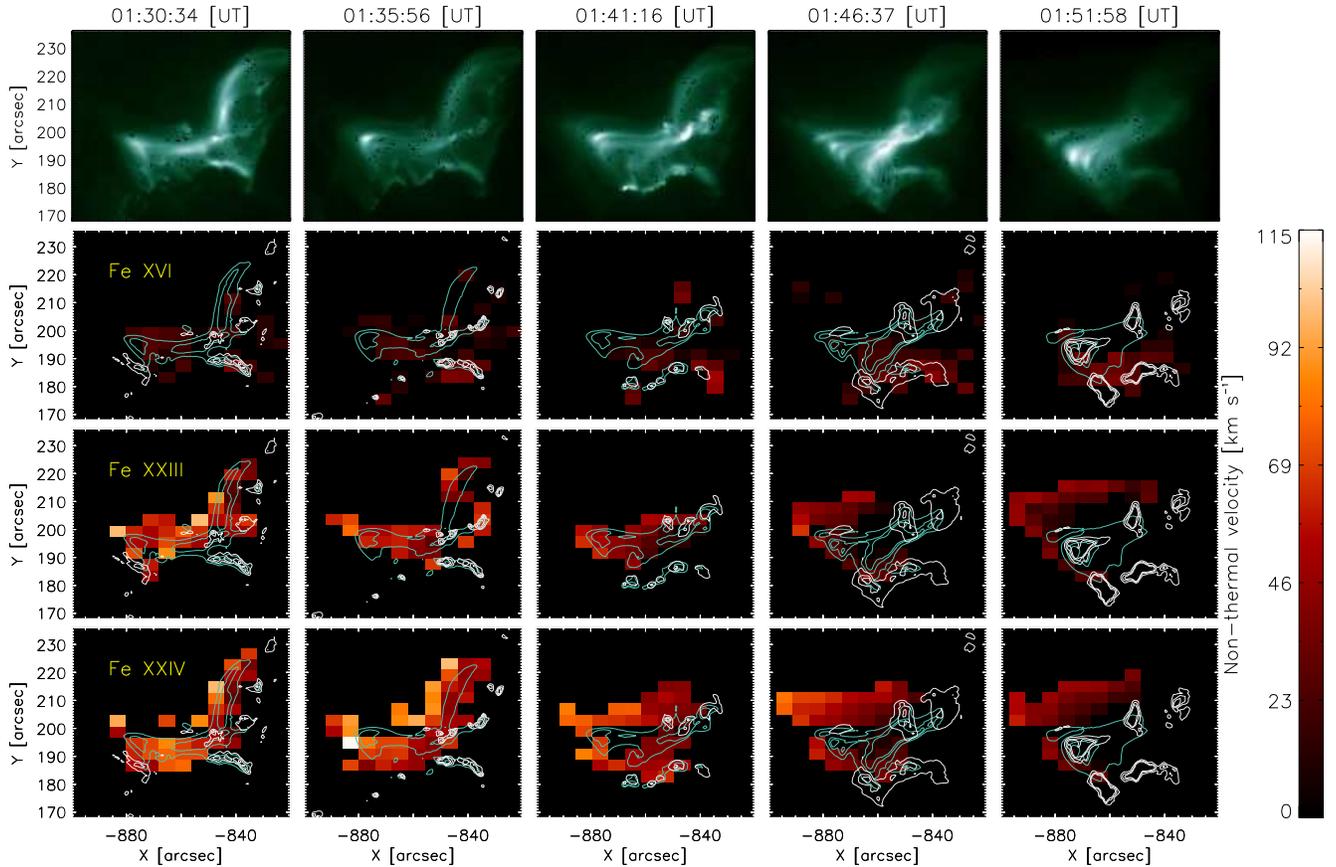}
    \caption{AIA 94 \AA~images (\textit{row 1}) and non-thermal velocity maps (\textit{rows 2-4}) for solar flare SOL2013-05-15T01:45 at five observation times. Rows 2-4 each study a different ion: Fe \textsc{xvi}, Fe \textsc{xxiii}, and Fe \textsc{xxiv}. AIA data is overlaid, blue contour lines show the AIA channel 94 \AA~intensity levels that range between 25$\%$ to 80$\%$ (of the maximum 94 \AA~image value) highlighting the coronal source and certain loop tops and white contours show the 304 \AA~intensity levels that range between 20$\%$ to 90$\%$ (of the maximum 304 \AA~image value) highlighting the flare ribbons.}
    \label{vel_maps}
\end{figure*}

Maps of non-thermal velocity, together with AIA 94 \AA~images (top row), are shown in Figure \ref{vel_maps}, for each ion studied at each observed time, from 01:30:34 UT to 01:51:58 UT. Rows 2-4 show a different ion, increasing in temperature. From top to bottom this corresponds to Fe \textsc{xvi}, Fe \textsc{xxiii}, and Fe \textsc{xxiv}. Each column shows a different observation time, with the time increasing from left to right. 
The AIA 94 \AA~\footnote{94 \AA~is chosen since it suffers from less saturation compared to 131 \AA.} 
contours (extracted intensity levels that range between 25$\%$ and 80$\%$ of the maximum 94 \AA~image value) overlaid onto the non-thermal velocity maps clearly denote the flare cusp and hot loops/loop tops (blue) while cooler material such as the flare ribbons are shown by AIA 304 \AA~contours (white) at levels ranging from 20\% to 90\% of the maximum AIA 304 \AA~image value. 

The non-thermal velocity maps of Fe \textsc{xxiii}, and Fe \textsc{xxiv} follow similar trends.
Firstly, we observe the presence of relatively high non-thermal velocities throughout the flare structures with values ranging between $v_{\rm nth}=113.24 \pm 2.40$ km s$^{-1}$ and $v_{\rm nth}=22.70 \pm 1.30$ km s$^{-1}$ early in the flare, before the peak in HXRs. 
At these early times, there is no strong pattern in space associated with the non-thermal velocity (or turbulence) and the $v_{\rm nth}$ values in these maps appear more random or `chaotic' compared to later times.
However, as the flare progresses to times during and after the HXR peak (01:41:16 UT to 01:51:58 UT), clear spatial patterns appear, with the largest values of non-thermal velocity present at the coronal loop top/cusp and decreasing towards the ribbon features/smaller radial distances in the map. 
By the last observation, at 01:51:58 UT, the region of largest non-thermal velocity is in the upper left quadrant of the maps (approximately $-900"\le x\le -870"$ and $200"\le y\le 215"$). Much of this region does not overlap with the AIA 94 \AA~contours in this map; these regions of high non-thermal velocity now sit at greater radial distances compared to the highest 94 \AA~intensity values in the lower loops.
At this time,  the cooler Fe \textsc{xvi} maps show non zero non-thermal velocity values closer to the lower loop region shown by the 94 \AA~contours, suggesting that the cooling plasma in these loops may still suffer from some turbulent plasma motions at decay times.

There is an EIS observation of the active region available prior to the start of the flare, at 00:45:02 UT. Non-thermal velocity maps at this non-flaring time (not shown in Figure \ref{vel_maps}) for the two hottest ions (Fe \textsc{xxiii} and Fe \textsc{xxiv}) are either non-existent or `noisy', as expected. However the cooler ion, Fe \textsc{xvi} already shows non-thermal broadening in the region where the solar flare occurs. 
During this pre-flare time, over the ranges of $-891.86''\le x \le -814.06''$ and $134.48''\le y \le 206.48''$, the space-averaged non-thermal velocity ($\langle v_{\rm nth} \rangle$)for Fe \textsc{xvi} is 17.82 $\pm 9.81$ km s$^{-1}$.

\begin{figure*}
    \centering
    \includegraphics[scale = 0.7]{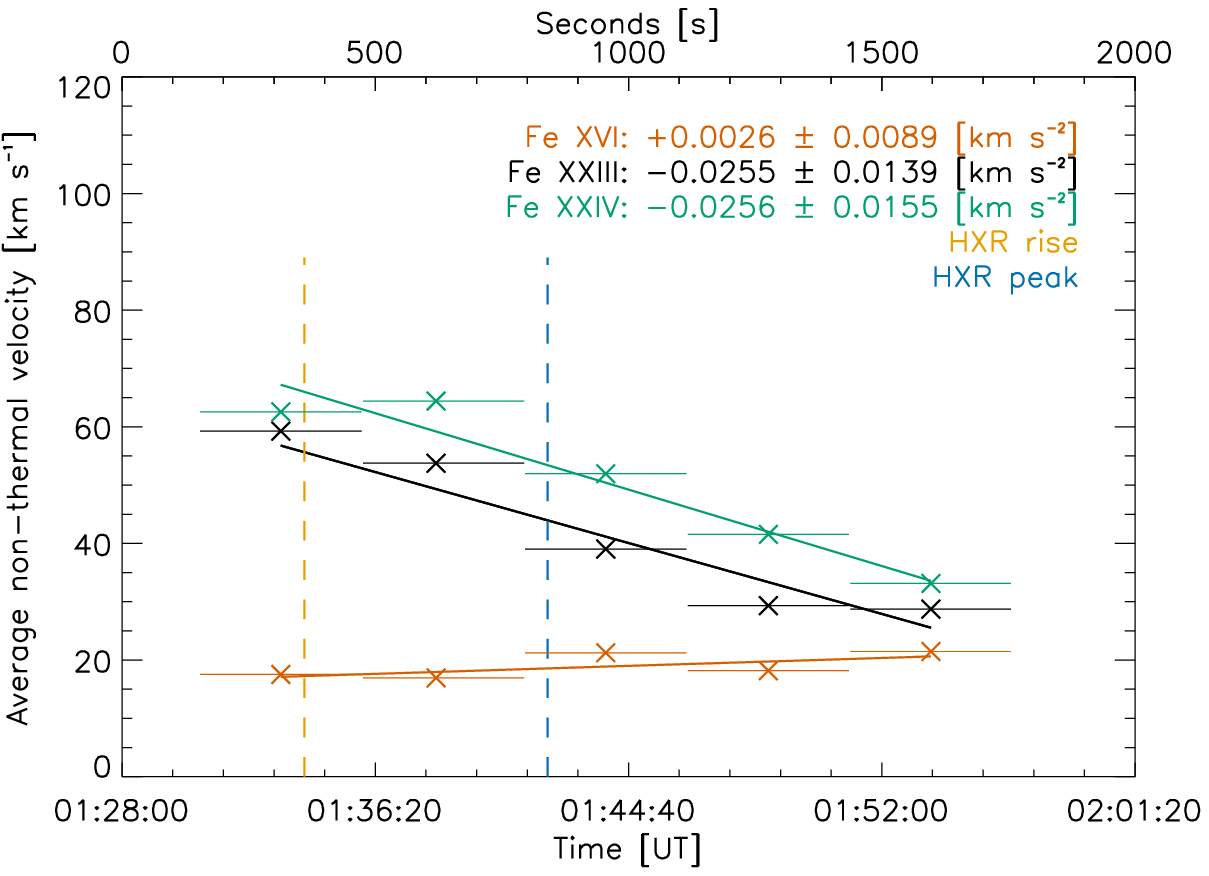}
    \includegraphics[scale = 0.7]{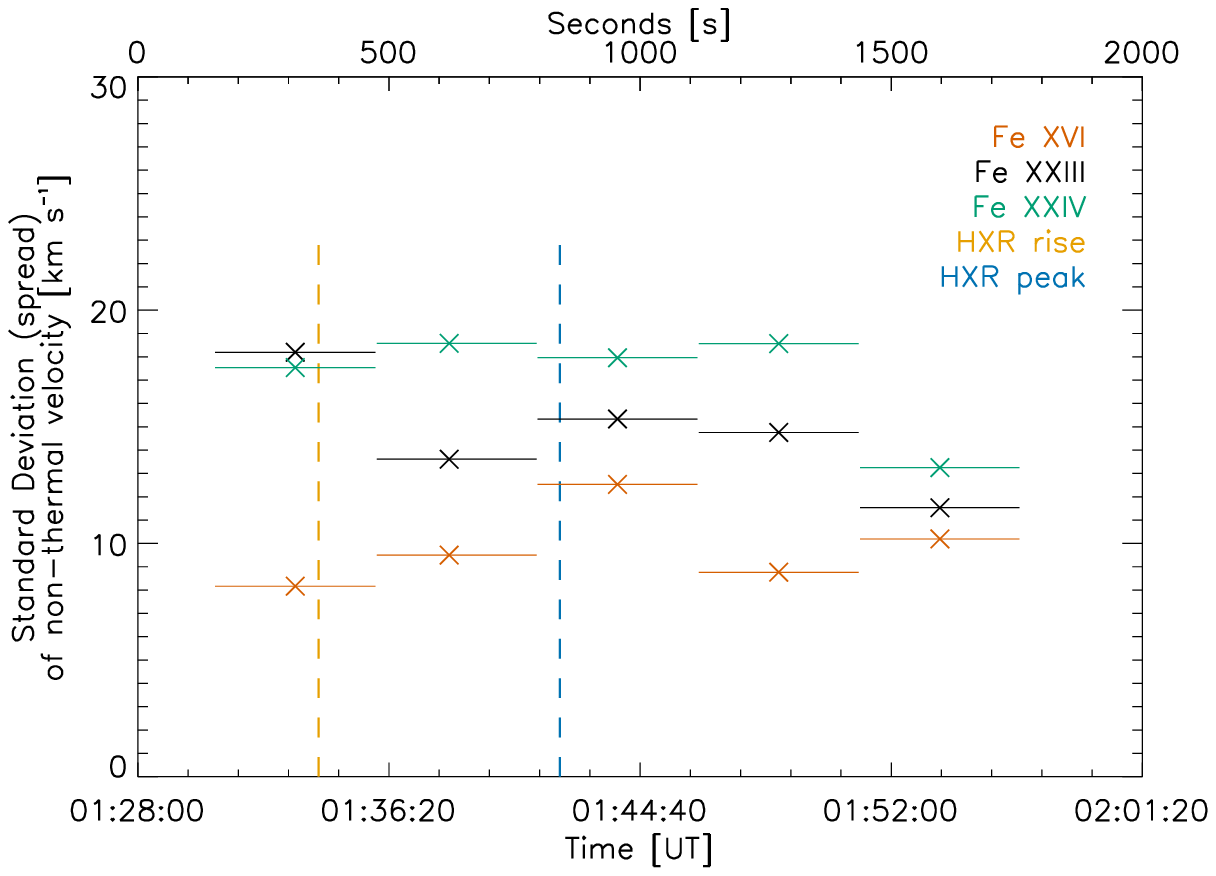}
    \caption{Space-averaged non-thermal velocity $\langle v_{\rm nth} \rangle$ (\textit{left}) and standard deviation (spread) of the non-thermal velocity values (\textit{right}) calculated from all pixels shown in each map in Figure \ref{vel_maps}, versus time for Fe \textsc{xvi} (orange), Fe \textsc{xxiii} (black), and Fe \textsc{xxiv} (teal). The data points are positioned in the center of the raster time. The approximate times of the HXR rise and peak are shown by the yellow and blue dashed lines in each panel. The values of $d\langle{v}_{\rm nth}\rangle/dt\;\pm$ uncertainty [km s$^{-2}$] are stated in the legend (left panel).} 
    \label{av_vel_time}
\end{figure*}

For the flare times, Figure \ref{av_vel_time} (left) shows the space-averaged non-thermal velocity calculated from all the values displayed in each map in Figure \ref{vel_maps}. The space-averaged non-thermal velocity $\langle v_{\rm nth} \rangle$ versus time for Fe \textsc{xvi}, Fe \textsc{xxiii}, and Fe \textsc{xxiv} is shown in orange, black and teal respectively. 
A linear fit is used to determine the rate at which $\langle v_{\rm nth} \rangle$ changes for each ion ($d\langle v_{\rm nth} \rangle/dt$), the value of which is stated in the Figure \ref{av_vel_time} legend.
Horizontal bars in Figure \ref{av_vel_time} show the raster time of the observation with the data points located in the center. The standard deviation (spread) of all $v_{\rm nth}$ values in each map is also shown in Figure \ref{av_vel_time} (right). Overall, the standard deviation (spread of $v_{\rm nth}$ values) in each map remains fairly constant for each ion as the flare progresses (over the rise, peak and decay) remaining at approximately 10 km s$^{-1}$, 15 km s$^{-1}$ and 18 km s$^{-1}$ for Fe \textsc{xvi}, Fe \textsc{xxiii} and Fe \textsc{xxiv} respectively. However at the last observation time, all ions have similar standard deviation values. The standard deviation (spread) values are used to calculate the uncertainties for $d\langle v_{\rm nth} \rangle/dt$. 

In general and as expected, the hotter ions, Fe \textsc{xxiii} and Fe \textsc{xxiv}, have greater values of space-averaged non-thermal velocity $\langle v_{\rm nth} \rangle$ compared to the cooler ion, Fe \textsc{xvi}. The space-averaged non-thermal velocity of the hotter ions Fe \textsc{xxiii} and Fe \textsc{xxiv} peaks at the beginning of the flare ($\langle v_{\rm nth} \rangle\approx 60-65$ km s$^{-1}$) and then decreases as the flare evolves, decreasing at a similar rate for both ions (to values between $\langle v_{\rm nth} \rangle\approx 30-35$ km s$^{-1}$), with a deceleration of $d\langle{v}_{\rm nth}\rangle/dt=-0.0255\pm0.0139$ km s$^{-2}$ and $d\langle{v}_{\rm nth}\rangle/dt=-0.0256\pm0.0155$ km s$^{-2}$ respectively.

Unlike the hotter ions, $d\langle v_{\rm nth} \rangle/dt$ for Fe \textsc{xvi} remains fairly constant. Furthermore, as previously mentioned, prior to the flare at 00:45:02 UT, Fe \textsc{xvi} shows a space-averaged non-thermal velocity of $\langle v_{\rm nth} \rangle=17.82\pm 9.81$ km$\,$s$^{-1}$, which is almost identical to the value of $\langle v_{\rm nth}\rangle$ at the onset of the flare of $\langle v_{\rm nth} \rangle=17.53\pm 8.16$ km$\,$s$^{-1}$. Thus, the overall space-averaged non-thermal velocity for Fe \textsc{xvi} does not change significantly during the flare. 

\subsection{Extracted map regions - temporal changes in $v_{\rm nth}$}
\begin{figure*}
    \centering
    \includegraphics[scale =0.8]{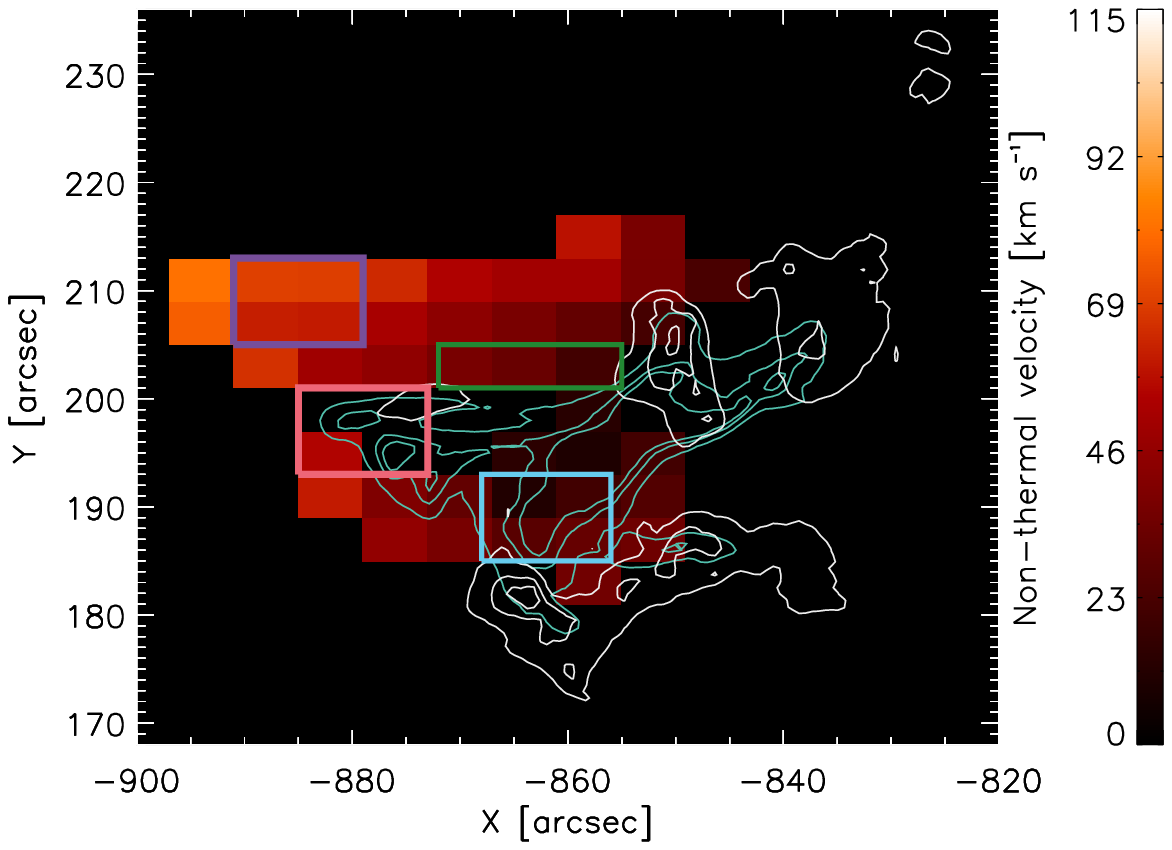}
    \includegraphics[scale = 0.7]{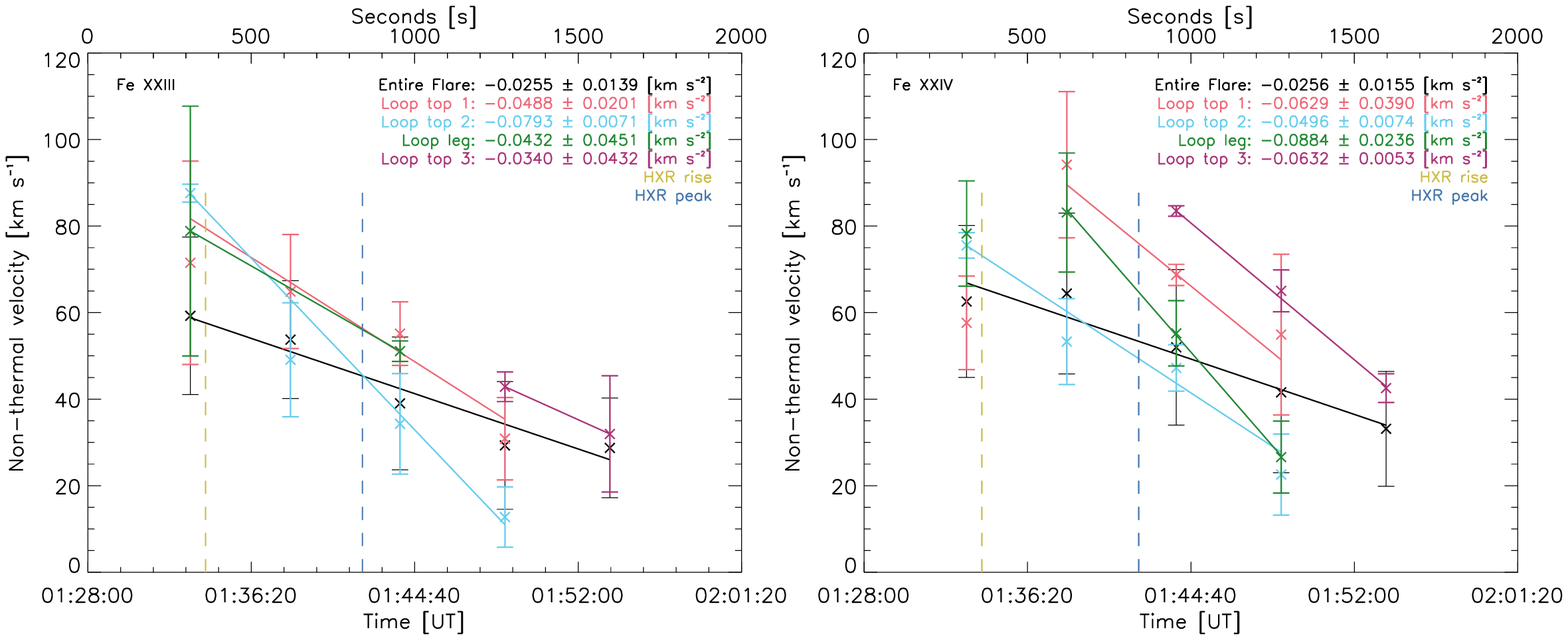}
    \caption{\textit{Top} Fe \textsc{xxiv} non-thermal velocity map at time 01:46:37 UT with the chosen extracted regions highlighted by the pink, purple, blue and green boxes. Pink and blue boxes cover coronal loop tops (named `loop top 1' and `loop top 2' respectively), the green box covers part of the loop leg connected to `loop top 1', and the purple box covers a region where bright pixels appear at later times (after the HXR peak and named `loop top 3'). Contours from AIA channel 94 \AA~with intensity levels that range between 25$\%$ to 80$\%$ of the maximum 94 \AA~image value (blue) and channel 304 \AA~with intensity levels ranging between 20$\%$ to 90$\%$ of the maximum 304 \AA~image value (white) are also plotted. \textit{Bottom} Non-thermal velocity against time for Fe \textsc{xxiii} (\textit{left}) and Fe \textsc{xxiv} (\textit{right}) for each extracted region (pink, blue, green and purple) and for all pixels (space-averaged $\langle v_{\rm nth} \rangle$) in the Figure \ref{vel_maps} map (black). The gradient$\pm$uncertainty ($dv_{\rm nth}/dt$ [km s$^{-2}$]) of each fit are shown in the legend. The approximate rise and peak in HXR emission is shown by the yellow and dark blue dashed lines in each panel, respectively. Raster time bars have been omitted from this figure for clarity; the data points are positioned in the center of the raster time.}
    \label{hot_extract}
\end{figure*}

Using the non-thermal velocity maps in Figure \ref{vel_maps}, Figure \ref{hot_extract} identifies several key features of the flare for further study (e.g., loop tops, loop leg) and shows how the Fe \textsc{xxiii} and Fe \textsc{xxiv} non-thermal velocities within these regions evolve in time, while Figure \ref{cool_extract} examines how the non-thermal velocity for Fe \textsc{xvi} evolves with time in one region close to the southern flare ribbon.
In each extracted region, the average non-thermal velocity (denoted simply as $v_{\rm nth}$ to distinguish it from the space-averaged value of the entire map $\langle v_{\rm nth} \rangle$) is calculated plus the standard deviation (spread of $v_{\rm nth}$ values in each region). For data points where only one pixel was available the uncertainty was determined using the uncertainty given for $\Delta\lambda$ by the Gaussian fit. Overall the non-thermal velocity within all of the chosen regions in Figure \ref{hot_extract} decrease as the flare evolves. We note that some data is missing in Figure \ref{hot_extract}, due to missing pixels at certain times.
\subsubsection{Coronal loop top regions}
In Figure \ref{vel_maps} (top row), the AIA 94 \AA~data shows multiple loops in the flare. Firstly, to study the non-thermal velocity in the different coronal loop tops in more detail, the pixels covering two identified loop top regions were extracted, highlighted by the pink and blue boxes in Figure \ref{hot_extract}. Here, `loop top 1' will refer to the pink box, which contains the loop top of the larger loop situated at the greater radial distance, and `loop top 2' will refer to the blue box containing the loop top of the smaller loop located at the smaller radial distance.

For both Fe \textsc{xxiii} and Fe \textsc{xxiv}, pixels are available for study in both regions `loop top 1' and `loop top 2' at all times except 01:51:58 UT. For Fe \textsc{xxiv}, at all times apart from the first studied (01:30:34 UT), the non-thermal velocity of `loop top 1' (pink) is greater than `loop top 2' (blue). We see a similar trend for Fe \textsc{xxiii}. 
For both ions and at all times, neglecting 01:30:34 UT, the non-thermal velocity in `loop top 1' is either greater than or approximately equal (at 01:46:37 UT) to the space-averaged non-thermal velocity of the entire map ($\langle v_{\rm nth}\rangle$).
For `loop top 2', the values of non-thermal velocity are greater at early times but quickly fall below the values of $\langle v_{\rm nth}\rangle$ seen in Figure \ref{av_vel_time}.

In each region, we investigate how the non-thermal velocity changes with time using linear fitting and by calculating the gradient $dv_{\rm nth}/dt$.
In `loop top 1' and `loop top 2' for both Fe \textsc{xxiii} and Fe \textsc{xxiv}, the decline in the non-thermal velocity ($dv_{\rm nth}/dt$) occurs at a faster rate than the decline in the space-averaged non-thermal velocity across the entire map ($d\langle{v}_{\rm nth}\rangle/dt$).
Overall, the analysis shows that $dv_{\rm nth}/dt$ varies in different regions across the maps, which might suggest: (1) the energy of the plasma motions is drained more efficiently in certain regions (e.g., for heating and/or particle acceleration) and/or (2) the mechanism producing the macroscopic motions diminishes quicker in certain regions.

In Figure \ref{hot_extract}, we also choose a region covering `loop top 3' (purple box). This covers the flare loops at the greatest radial distance, mainly produced at later times in the flare, after the peak in HXRs. 
Located above both `loop top 1' and `loop top 2', this region is not observed in the AIA 94~\AA~ contours ranging between 25\% and 80\% of the maximum 94 \AA~image value. 
For the data points available, the non-thermal velocity in `loop top 3' also decreases as the flare progresses. Comparing each loop top region shows that at a given time (when available), the largest values of $v_{\rm nth}$ occur at the greater radial distance so that $v_{\rm nth}$ (`loop top 3') $>$  $v_{\rm nth}$ (`loop top 1') $>$ $v_{\rm nth}$ (`loop top 2').

\subsubsection{Loop leg}
In Figure \ref{hot_extract}, part of the loop leg that appears to be connected to `loop top 1' is highlighted by the green box (this will be referred to as the `loop leg').  As with `loop top 1', the Fe \textsc{xxiv} non-thermal velocity in `loop leg' only decreases with time after 01:35:56 UT, once the HXR emission begins to rise. Hence, for Fe \textsc{xxiv} we only consider observation times from 01:35:56 UT, where $v_{\rm nth}$ decreases with time when studying the gradient ($dv_{\rm nth}/dt$) in this region. For Fe \textsc{xxiv}, from 01:35:56 UT onward, `loop leg' $v_{\rm nth}$ values are smaller than `loop top 1' $v_{\rm nth}$ values and they decrease quicker over the same time period compared to `loop top 1' (with gradient $dv_{\rm nth}/dt=-0.0884\pm0.0236$ km s$^{-1}$ compared to $dv_{\rm nth}/dt=-0.0629\pm0.0390$ km s$^{-1}$, respectively). For Fe \textsc{xxiii}, only times of 01:30:34 UT and 01:41:16 UT are available for study. For Fe \textsc{xxiii}, the values of `loop leg' $v_{\rm nth}$ are similar to `loop top 1' values and $v_{\rm nth}$ decreases at approximately the same rate in both regions (with gradient $dv_{\rm nth}/dt=-0.0432\pm0.0451$ km s$^{-1}$ compared to $dv_{\rm nth}/dt=-0.0488\pm0.0201$ km s$^{-1}$, respectively).

\subsubsection{Southern ribbon}
\begin{figure*}
    \centering
    \includegraphics[width = 0.49 \linewidth]{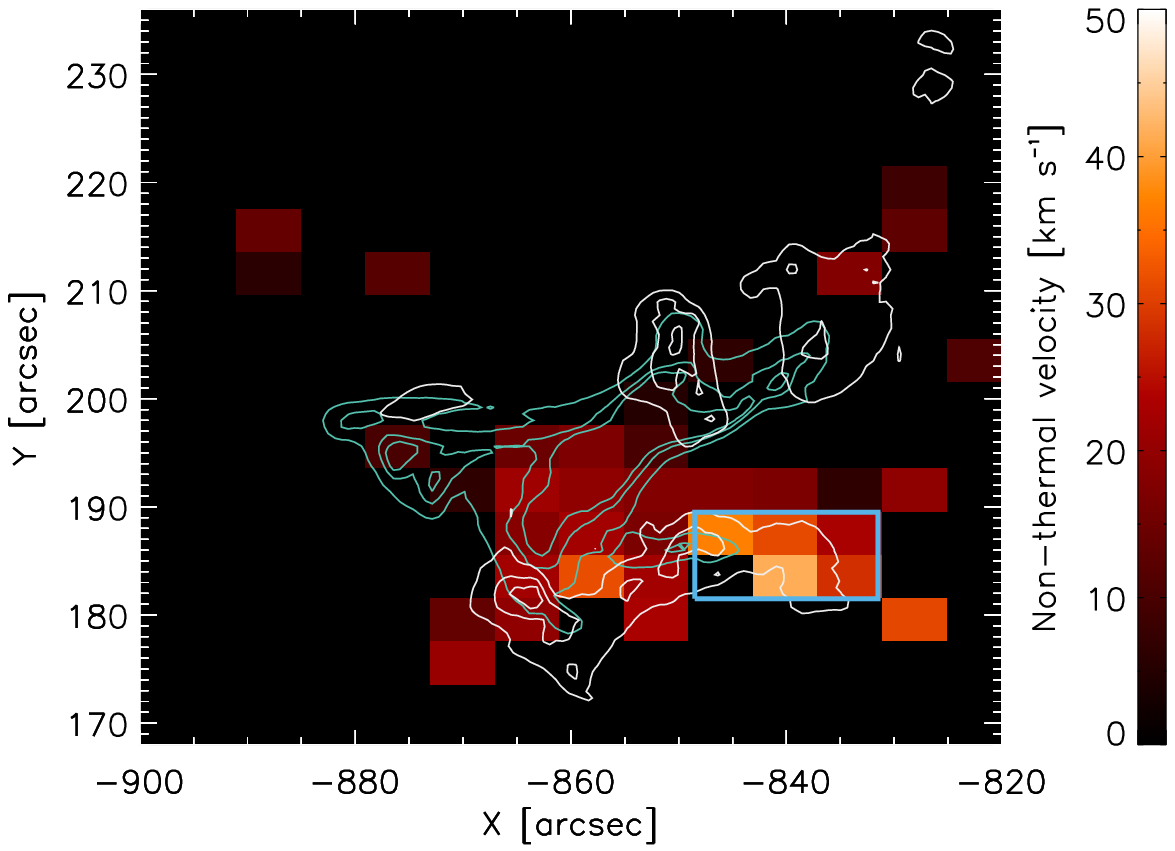} 
    \includegraphics[width = 0.49 \linewidth]{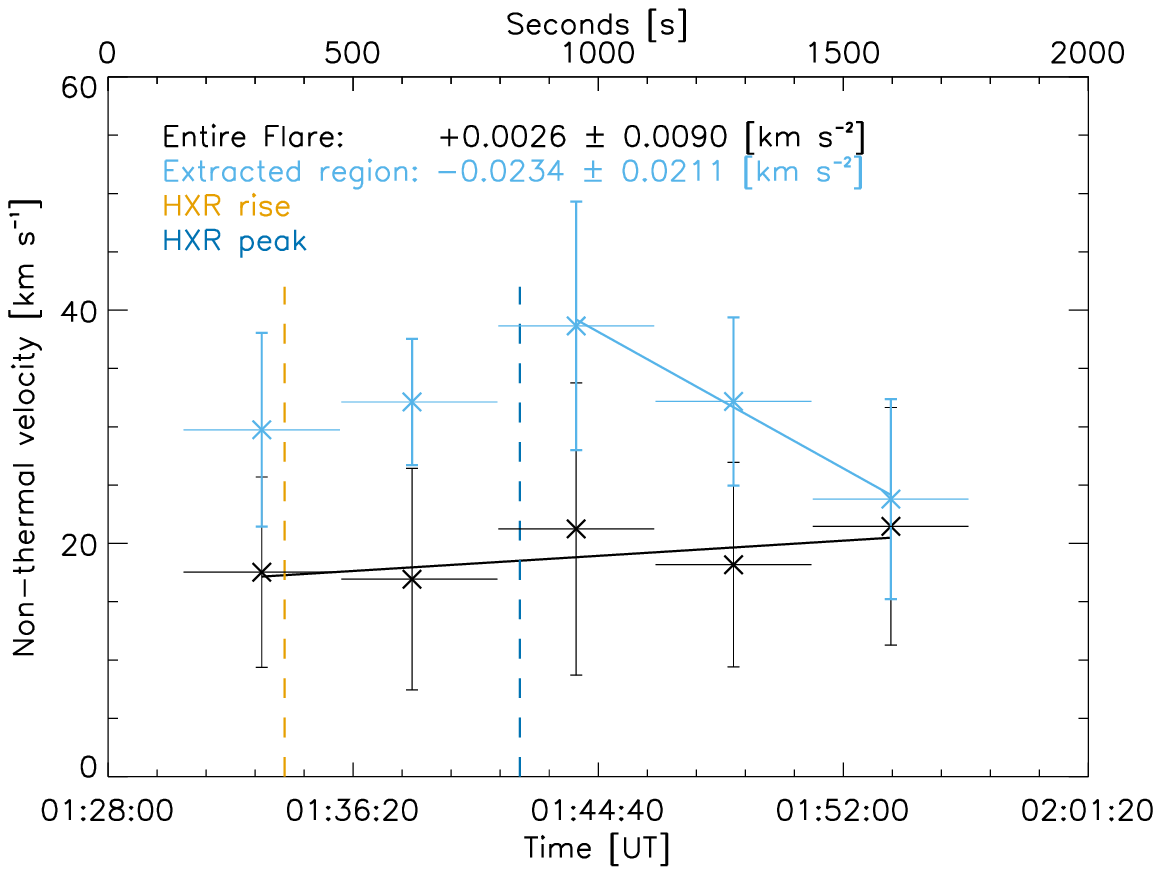}
    \caption{Non-thermal velocity map (\textit{left}) of Fe \textsc{xvi} at time 01:46:37 UT with AIA 304 \AA~contours ranging between 20$\%$ to 90$\%$ of the maximum 304 \AA~image value (white) and 94 \AA~contours ranging between 25$\%$ to 80$\%$ of the maximum 94 \AA~image value (blue). The blue box shows the area extracted. Fe \textsc{xvi} non-thermal velocity against time (\textit{right}) for the extracted region (blue) and for all pixels (space-averaged $\langle v_{\rm vth} \rangle$) in the Figure \ref{vel_maps} map (black). The gradient$\pm$uncertainty ($dv_{\rm nth}/dt$ [km s$^{-2}$]) of each linear fit are shown in the legend. The approximate rise and peak times of HXR emission are shown by the yellow and dark blue dashed line respectively. }
    \label{cool_extract}
\end{figure*}

We also study how the non-thermal velocity from the cooler ion Fe \textsc{xvi} changes with time in one region, shown by the blue box in Figure \ref{cool_extract}, where the Fe \textsc{xvi} non-thermal velocity appears to increase along the lower edge of the flare. The AIA 304 \AA~data shows that this area corresponds to the southern flare ribbon on the map (although the exact height of the Fe \textsc{xvi} emission is unknown). 
Extracting the average $v_{\rm nth}$ in this region and comparing it to the space-averaged value of the entire flare $\langle v_{\rm nth} \rangle$ shows that $v_{\rm nth}$ in this area is higher than $\langle v_{\rm nth} \rangle$ at each studied time. 
For example, at 01:35:56 UT $\langle v_{\rm nth} \rangle =$ 16.93 $\pm 9.50$ km$\,$s$^{-1}$ but $v_{\rm nth} =$ 32.13 $\pm$ 5.41 km$\,$s$^{-1}$ in the extracted blue box region.

Initially, the non-thermal velocity of the extracted region increases, coinciding with the rise in HXRs which begins at 01:34:00 UT (indicated by the yellow dashed line in Figure \ref{cool_extract}).
The non-thermal velocity peaks at 01:41:16 UT with a value of $v_{\rm nth} =$ 38.65 $\pm$ 10.66 km$\,$s$^{-1}$ and at approximately the same time as the peak in HXRs at 01:42:00 UT, as shown by the blue dashed line in Figure \ref{cool_extract}.
After the peak in HXRs, the non-thermal velocity of the extracted region begins to decrease. By the last observation time, at 01:51:58 UT, the value of the non-thermal velocity $v_{\rm nth}$ is approximately the same value as the space-averaged non-thermal velocity of the entire flare $\langle v_{\rm nth} \rangle$.

\subsection{Spatial changes in $v_{\rm nth}$}
\label{Sect:spaital}

\begin{figure*}
    \centering
    \includegraphics[scale = 0.8]{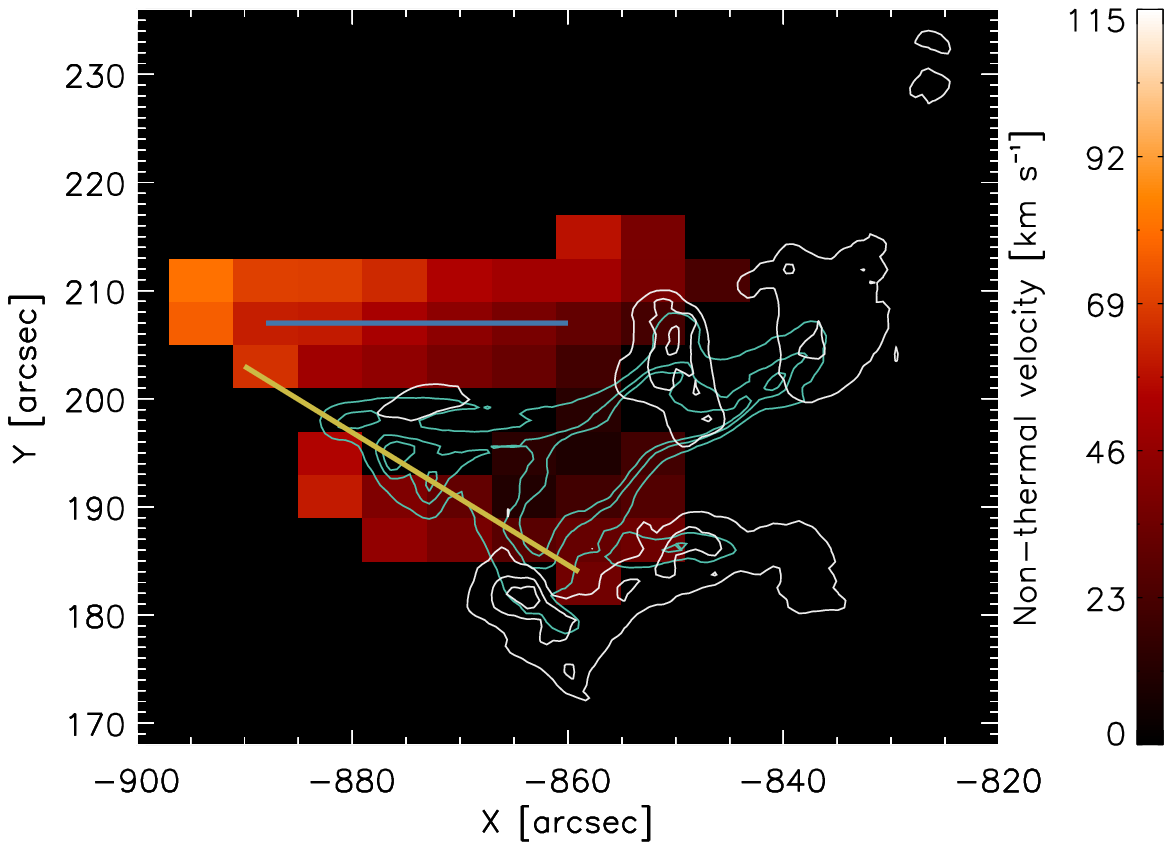}
    \includegraphics
    [scale = 0.7]{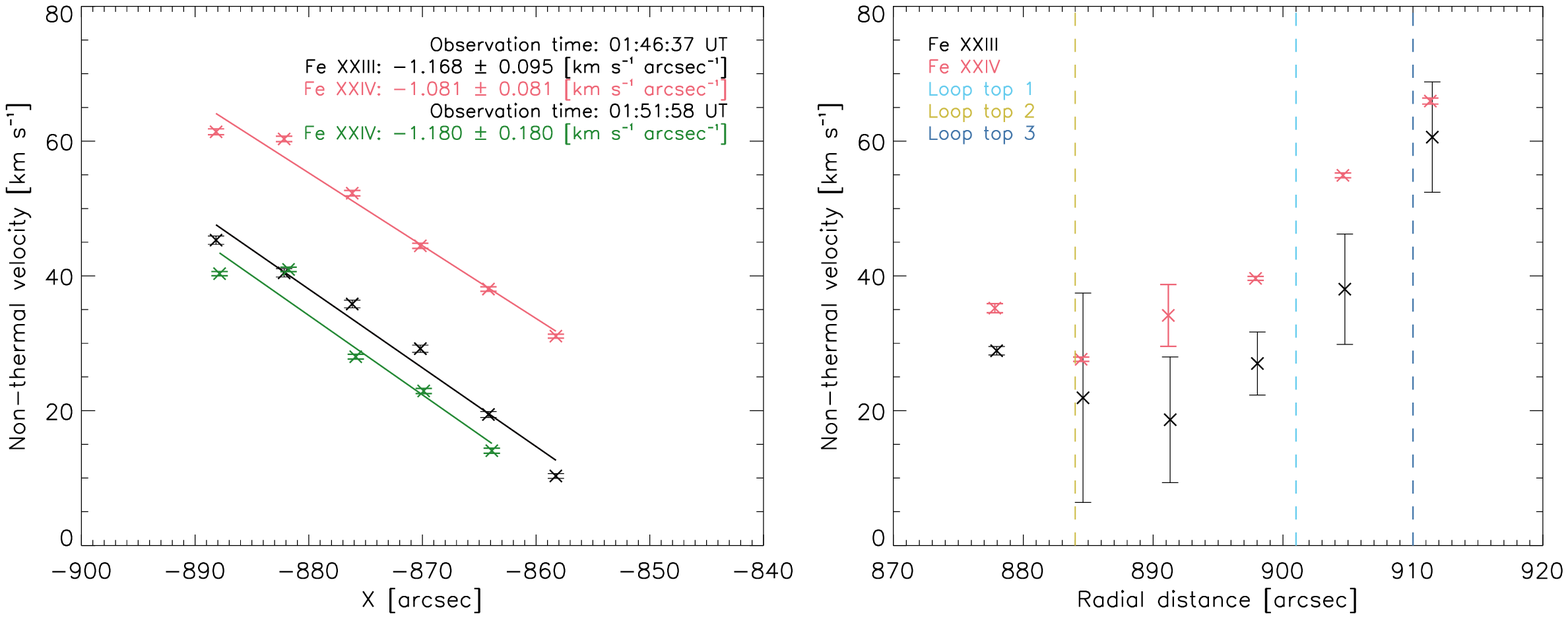}
    \caption{Non-thermal velocity map (\textit{top}) of Fe \textsc{xxiv} at 01:46:37 UT. Changes in non-thermal velocity with space are examined along the line joining the coronal loop top source, loop leg and northern ribbon (blue line) and along the line connecting the upper and lower loop top sources (yellow line). \textit{Bottom left:} The non-thermal velocity versus solar $x$ (along the blue line) for Fe \textsc{xxiii} and Fe \textsc{xxiv} at 01:46:37 UT (black and pink respectively) and for Fe \textsc{xxiv} at 01:51:58 UT (green). The gradient$\pm$uncertainty ($dv_{\rm nth}/dx$ [km s$^{-1}$ arcsec$^{-1}$]) of each linear fit are shown in the legend. \textit{Bottom right}: The non-thermal velocity versus radial distance $r=\sqrt{x^{2} + y^{2}}$ (along the yellow line) for Fe \textsc{xxiii} and Fe \textsc{xxiv} at 01:46:37 UT (black and pink respectively) The light blue, yellow, and dark blue dashed lines show the approximate location of `loop top 1', `loop top 2', and `loop top 3' respectively.}
    \label{space_grad}
\end{figure*}

In the maps of non-thermal velocity in Figure \ref{vel_maps}, $v_{\rm nth}$ changes in space with spatial patterns becoming clearer during and after the peak in HXRs, particularly for the hotter ions Fe \textsc{xxiii} and Fe \textsc{xxiv}. Here we focus on the observation at 01:46:37 UT, where spatial patterns in $v_{\rm nth}$ are clearly seen for both Fe \textsc{xxiii} and Fe \textsc{xxiv} (although similar patterns can be observed during the time close to the peak in HXRs at 01:41:16 UT and at the later time of 01:51:58 UT). 
We consider changes in space along the two lines shown in Figure \ref{space_grad}: the horizontal blue line covers the region from a loop apex (previously `loop top 3') and along the upper loop leg; the diagonal yellow line passes through the middle of the flare coronal source and multiple coronal loop tops.
There are no Fe \textsc{xvi} pixels close to `loop top 3', thus only the hotter ions are considered in Figure \ref{space_grad}. It was difficult to study spatial changes at earlier times during the rise in HXRs. This was due to the compactness of the loop tops and the more chaotic pattern of non-thermal velocity at these times, suggesting that the spatial distribution of turbulence may be more random and possibly dependent on fluctuations in local conditions at early times in the flare.

For each line, we plot the non-thermal velocity against position ($v_{\rm nth}$ against $x$ for the blue line and against radial distance $r=\sqrt{x^{2}+y^{2}}$ for the diagonal yellow line), and calculate the gradient $dv_{\rm nth}/dx$ from the values determined along the blue line.
Along the blue line, we also consider changes in non-thermal velocity versus $x$ at the later time of 01:51:58 UT for Fe \textsc{xxiv}, shown in green in Figure \ref{space_grad}. Due to insufficient pixels, Fe \textsc{xxiii} was not included on the plot at this time. Along the blue line, only single pixels from Figure \ref{vel_maps} are considered.

Along the blue line at time 01:46:37 UT, for both Fe \textsc{xxiii} and Fe \textsc{xxiv}, the values of $v_{\rm nth}$ fall as we move from the loop apex along the loop leg. 
The non-thermal velocity decreases at a similar rate for both Fe \textsc{xxiii} and Fe \textsc{xxiv} with $dv_{\rm nth}/dx=-1.168 \pm 0.095$ km s$^{-1}$ arcsec$^{-1} $ and $dv_{\rm nth}/dx=-1.081 \pm 0.081$ km s$^{-1}$ arcsec$^{-1}$, respectively. Along the blue line at 01:51:58 UT, the values of non-thermal velocity for Fe \textsc{xxiv} are approximately $\approx$ 20-25 km s$^{-1}$ lower than at the previous time but the values decrease along $x$ at roughly the same rate of $dv_{\rm nth}/dx=-1.180 \pm 0.180$ km s$^{-1}$ arcsec$^{-1} $. Along the blue line at 01:51:58 UT, the values of non-thermal velocity for Fe \textsc{xxiv} are similar to the values for Fe \textsc{xxiii} at 01:46:37 UT. 

As discussed, the AIA 94~\AA~maps (top row in Figure \ref{vel_maps}) show multiple flare loops (studied using `loop top 1', `loop top 2' and `loop top 3' in Figure \ref{hot_extract}). The diagonal yellow line in Figure \ref{space_grad} cuts through these different loop tops. To determine the non-thermal velocity along this line, the average of 2 pixels covering the line is used. The approximate locations of `loop top 1', `loop top 2' and `loop top 3' are highlighted in Figure \ref{space_grad} (bottom right) by the light blue, yellow, and dark blue dashed lines respectively. For both Fe \textsc{xxiii} and Fe \textsc{xxiv}, the non-thermal velocity decreases overall from the larger loops at the greater radial distance to smaller loops at a smaller radial distance. For Fe \textsc{xxiv}, from `loop top 3' to `loop top 1' the non-thermal velocity decreases at a faster rate than from `loop top 1' to `loop top 2'. At the radial distances smaller than `loop top 2' the non-thermal velocity increases for both ions.

It is interesting to note that the fall in non-thermal velocity (turbulence) is approximately linear along the blue line connecting the loop apex and loop leg at times corresponding to the peak in HXRs and after (and approximately linear along the yellow radial line connecting the different loop tops). In recent modeling studying the effects of an extended spatial distribution of turbulence on electron acceleration e.g., \citet{2018A&A...612A..64S}, the turbulent acceleration diffusion coefficient is modeled as a decreasing exponential in space (along the coronal loop) since the form of the spatial distribution of turbulence accelerating electrons in the flare is usually unknown. However, although the exact relationship between the diffusion coefficient and the macroscopic velocity $v_{\rm nth}$ is not entirely clear, we can see that the decrease in $v_{\rm nth}$ in space along the loop is much slower than exponential, suggesting that the turbulence may influence the acceleration or transport of electrons over a larger spatial extent than modeled to date. Such observations are extremely useful for constraining future simulations studying the link between spatially distributed turbulence and electron acceleration and transport in flares.

\subsection{Turbulent kinetic energy density maps}

\begin{figure*}
    \centering
    \includegraphics[scale = 0.74]{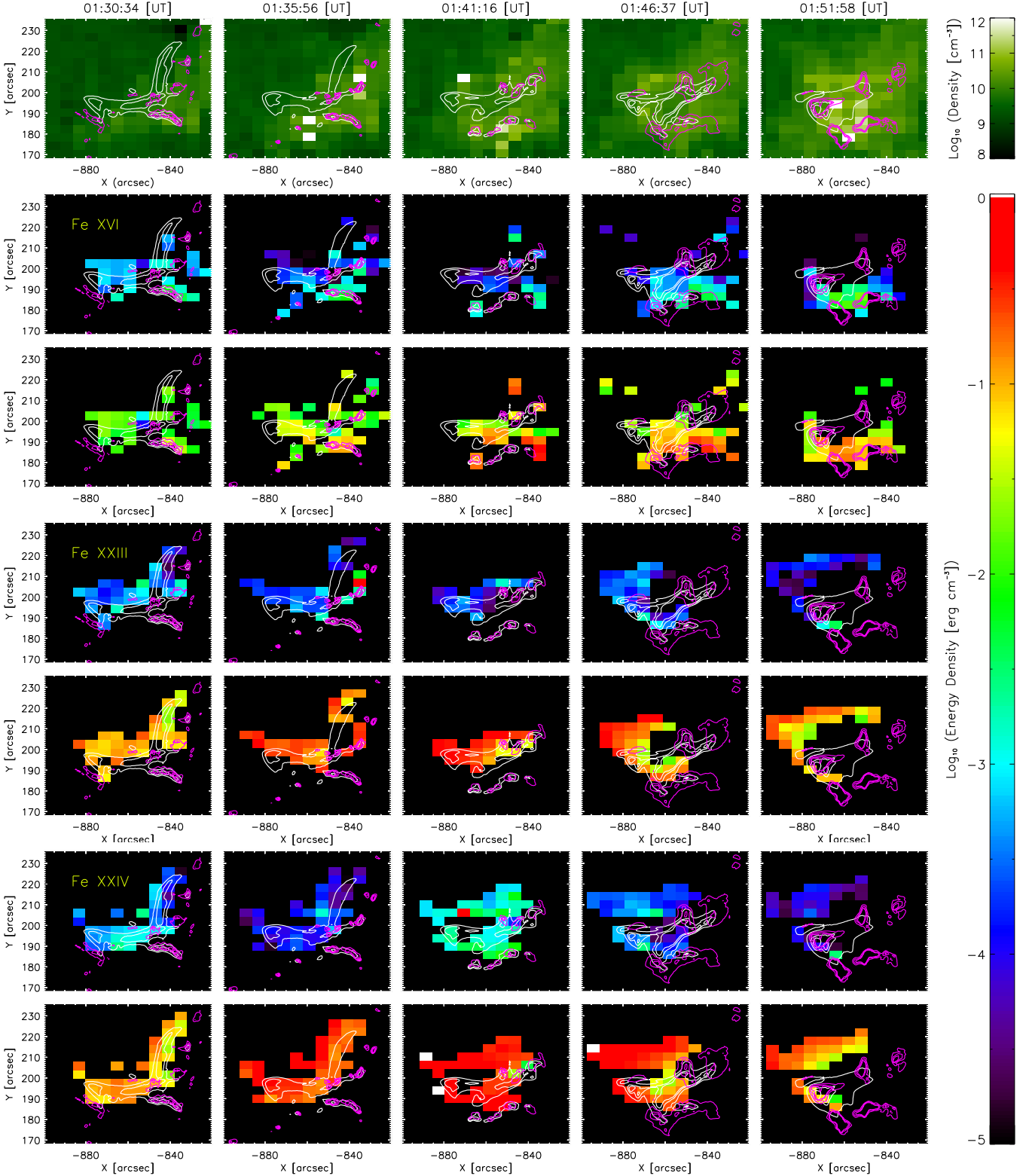}
    \caption{Density ($n_{e}$ [cm$^{-3}$]) and kinetic energy density ($K/V$ [erg cm$^{-3}$]) maps of the flare (time increasing from left to right) overlaid with AIA 94 \AA~contours (ranging between 35$\%$ to 70$\%$ of the maximum 94 \AA~value: white) and 304 \AA~contours (ranging between 20$\%$ to 90$\%$ of the maximum 304 \AA~value: pink). \textit{Top row}: ($\log_{10}$) density maps for each observation time created using the Fe \textsc{xiv} 264.79 \AA/274.20 \AA~intensity ratio. The second and third rows show the ($\log_{10}$) kinetic energy density maps for Fe \textsc{xvi} using the spatially varying EIS density ratio (row 2) and single value space-averaged coronal density determined from X-ray observations in \citet{2017PhRvL.118o5101K} (row 3). Rows 4 and 5 show the same results (as row 2 and 3) but for Fe \textsc{xxiii}, while rows 6 and 7 show the same results but for Fe \textsc{xxiv}. For $\log_{10}(K/V)$, the background level (no values of $v_{\rm nth}$ or $K/V$) is set to a value 10$^{-5}$ for plotting purposes only.}
    \label{kinetic_maps}
\end{figure*}

In \citet{2017PhRvL.118o5101K}, a space-averaged value of non-thermal velocity $\langle v_{\rm nth} \rangle$ in the flaring corona of SOL2013-05-15T01:45 is converted to a space-averaged coronal kinetic energy $\langle K \rangle$ [erg] using 
\begin{equation}
\langle K \rangle=\frac{3}{2}\cdot\,1.3 m_{p} \langle v_{\rm nth}\rangle ^{2} n_{p} V
\end{equation}
where $m_{p}$ is the proton mass [g] (and $1.3\;m_{p}$ is the mean ion mass for solar coronal abundances \citep{2014SoPh..289..977R}), $n_{p}$ is proton number density [cm$^{-3}$] (taken as a single space-averaged coronal value at each time in \citet{2017PhRvL.118o5101K}) and $V$ is the plasma volume [cm$^{3}$]. 

For the first time, in order to calculate the kinetic energy $K$ at every spatial pixel shown in the non-thermal velocity maps (Figure \ref{vel_maps}), we employ spectral line density diagnostics, specifically using the Fe \textsc{xiv} (264.7889 \AA/~274.2037\AA) intensity ratio \citep{young2011deriving}. This method provides the electron number density $n_{e}$ in regions of the plasma where Fe \textsc{xiv} is produced at $\approx2$~MK. Making the assumption that $n_{e}=n_{p}$, $n_{p}$ can then be determined at every spatial pixel across the flare at each observation time, as shown in the top panel of Figure \ref{kinetic_maps}. The obvious issue with density determination using this Fe \textsc{xiv} intensity ratio, is that it might not provide a true indication of the density within the hotter flaring coronal loops where Fe \textsc{xiv} is not present, particularly during the rise and peak of the flare. However, we stress it is the only suitable density diagnostic available for the flare observation. We also note that Fe \textsc{xiv} (274.2037~\AA) contains a cool blend at 274.1804~\AA~formed at $\approx0.63$~MK. Density maps were also created accounting for this blend using the automatic EIS fitting software and we did not see any noticeable differences in the density maps produced when this blend was accounted for in the fitting procedure.

The density maps (row 1 of Figure \ref{kinetic_maps}) show that initially there is an increase in plasma density where the flare is located, shown by the AIA 94 \AA~contours, compared to the surrounding solar atmosphere. As the flare progresses, the area around the flare increases in density until almost the entire field of view of the map has a higher density than the background density observed in the first map. As with the density surrounding the flare, the density of the area covering the AIA 94 \AA~contours increases as the flare evolves. At 01:51:58 UT the area of highest density corresponds to a region where no values of non-thermal velocity are observed for Fe \textsc{xxiii} and Fe \textsc{xxiv}.

The turbulent kinetic energy density ($K/V$ [erg cm$^{-3}$]) was determined for each ion at each observation time, shown in rows 2,4 and 6 of Figure \ref{kinetic_maps} for Fe \textsc{xvi}, Fe \textsc{xxiii}, Fe \textsc{xxiv} respectively using the spatially resolved Fe \textsc{xiv} density maps and the non-thermal velocity maps shown in Figure \ref{vel_maps}. For clarity, the maps in Figure \ref{kinetic_maps} show $\log_{10}(K/V)$. For each ion studied in Figure \ref{kinetic_maps}, we also show the kinetic energy density ($K/V$) maps using the single space-averaged values of coronal number density calculated from \emph{RHESSI} X-ray spectroscopy and taken from \cite{2017PhRvL.118o5101K} (rows 3, 5, and 7). Using a single space-averaged value of coronal number density from X-ray spectroscopy should provide an approximate upper limit on both the number and kinetic energy density in each region of the flare. Using the space-averaged coronal density produces a map of kinetic energy density that only depends on the variations in non-thermal velocity and hence, similar patterns emerge spatially in these ($K/V$) maps and the non-thermal velocity maps in Figure \ref{vel_maps}. Interestingly, when the spatial changes in density are taken into account, the patterns observed in the non-thermal velocity maps (particularly where higher values of non-thermal velocity are seen at greater radial heights in Section \ref{Sect:spaital}) are diminished (compared to the constant single density case) particularly along the coronal loop apex radial line. Moreover, it was determined in \citet{jeffrey2015high} that the coronal loop top source of the flare can contain strong gradients in both temperature and number density along the radial direction opposite to the guiding magnetic field, varying by an order of magnitude with the density decreasing, and temperature increasing, towards greater radial heights.

Taking into account changes in density alongside changes in $v_{\rm nth}$ allows us to see important local inhomogeneities in the spatial distribution of the kinetic energy available. Moreover, even with the present of local fluctuations, the results are highly suggestive that an almost identical reservoir of kinetic energy from turbulent plasma motions is available in large parts of the flaring corona at a given time, indicating that this energy could be extracted by particles in the form of acceleration and heating in multiple parts of the flare. Away from the coronal loop top, we also see higher values of the kinetic energy density in cooler regions denoted by the Fe \textsc{xvi} emission. In regions closer to the flare ribbons, where the Fe \textsc{xvi} non-thermal velocity was shown to increase before the peak in HXR emission, the larger number densities produce large values of kinetic energy density in the region. However, the efficiency of particle acceleration will depend on the local plasma conditions and whether acceleration can overcome Coulomb collisions (local density), possibly leading to increased heating instead in very dense regions.

\section{Summary of the observations}
\label{sec:discussion}

The non-thermal broadening of spectral lines is a key sign of macroscopic plasma motions leading to turbulence in flares. In this paper, we mapped the non-thermal broadening (velocity) across a large region in a solar flare at different times and found the following main observations:

\begin{itemize}
    \item As is often observed, but not discussed, the presence of non-thermal broadening in flares is not a localized phenomenon. The maps in Figure \ref{vel_maps} show the presence of non-thermal velocities (turbulence) across the entire flare region covering the loop apex, loop legs and close to ribbon features lower in the atmosphere\footnote{Although the non-thermal broadening is aligned with features here, the exact height of turbulence in the atmosphere is not known from this data.}.
    \item Indeed as shown by e.g., \citet{1995ApJ...438..480A}, \citet{2014ApJ...788...26D}, non-thermal broadening (turbulence) is greatest for the hottest ions, Fe \textsc{xxiii} and Fe \textsc{xxiv}, and in general reduces with ion temperature, again suggesting that there is a temperature gradient to turbulence in plasma. Further, the MHD modeling of \citet{2016A&A...589A.104G} suggests that higher non-thermal velocities (turbulence) lead to greater heating and/or acceleration (particle energization) and hence higher ion temperatures, since more energy is dissipated. 
    \item Turbulence is not localized but the hottest ions available for study, Fe \textsc{xxiii} and Fe \textsc{xxiv}, show that the non-thermal broadening is often greatest in regions that coincide to the loop tops of magnetic loops, where it is expected (in a standard model) that the primary energy release and the bulk of electron energization occur.
    \item Considering the two hottest ions, Fe \textsc{xxiii} and Fe \textsc{xxiv}, Figure \ref{av_vel_time} shows that turbulence is greatest at the start of the flare and reduces over time as the mechanism producing turbulence dies away as expected. Before the peak in HXRs, the maps display a more random distribution of non-thermal velocities across the flare, but during and after the peak we see a clear spatial pattern with the largest non-thermal velocities occurring in the loop apex and decreasing towards the positions of the flare ribbons. The spatial variation in non-thermal velocity from a loop apex along the loop leg (and along the radial line connecting the different loop tops) appears to reduce at approximately a linear rate.
    \item At the beginning of the flare, the non-thermal velocity in the coronal source is higher than the average non-thermal velocity across the entire flare. Turbulence (non-thermal velocity) then reduces in this region at a faster rate than the flare overall, until the non-thermal velocity in the coronal region is similar to the value of the average non-thermal velocity in the entire flare. This might suggest that energy from turbulence in the coronal source is dissipated at a higher rate than in the rest of the flare. Alternatively, the mechanism producing turbulence may die faster in this region. In turbulence theory, the energy deposition rate indeed increases with $v_{\rm nth}$ since energy per mass unit/time $\sim = v_{\rm nth}^3/L_\perp$ where $L_\perp$ is the perpendicular scale of turbulence \citep[e.g., ][]{1995ApJ...438..763G}. In general, we find different values of $dv_{\rm nth}/dt$ in different regions of the corona.
    \item Turbulence is present in cooler, lower heights in the flare, i.e. among the flare ribbons. This is suggested by the increase and peak in non-thermal velocity in the extracted region of Fe \textsc{xvi} in Figure \ref{cool_extract}, which coincides with the rise and peak in HXRs as the flare progresses. After the peak in HXRs, the region continues to have an increased amount of turbulence which decreases until it reaches a similar level to the average value of Fe \textsc{xvi} non-thermal velocity in the entire flare. We note that the EIS time resolution cannot determine if turbulence appears before or affects the energetic electrons in the region or if the turbulence is generated by the energetic electrons reaching the lower atmosphere.
    \item When converted to a turbulent kinetic energy density, the use of a more realistic spatially-varying number density shows that although the non-thermal velocities decrease as we move away from the loop apex, changing densities away from the apex diminish the variations (or at least large gradients) in kinetic energy at different points in the flare suggesting that similar levels of kinetic energy may be available in regions away from the loop apex and into coronal loop legs and the lower corona (although the efficiency of particle energization in denser regions may be reduced).
\end{itemize}

This detailed study suggests that turbulence plays a role throughout the flaring corona and has a complex inhomogeneous spatial and temporal structure and evolution, not considered in previous studies. We await the availability of hot density diagnostics for plasma $>10$~MK from future instrumentation which will greatly improve the determination of the kinetic energy (density) in flaring coronal loops, helping to uncover both interesting local fluctuations and constrain the availability of kinetic energy in the flaring corona. This work should be of interest to the modeling community and in particular, help to constrain both the mechanism of turbulence in flares and its spatial and temporal distributions which are of great importance to the production and form of accelerated electrons in flares. 

\section*{acknowledgments}
NLSJ gratefully acknowledges the current financial support from the Science and Technology Facilities Council (STFC) Grant ST/V000764/1 and previous support from STFC Grant ST/P000533/1. MS gratefully acknowledges the financial support from the Northumbria University RDF studentship. The authors acknowledge IDL support provided by STFC. The work is supported by an international team grant \href{http://www.issibern.ch/teams/solflareconnectsolenerg/}{“Solar flare acceleration signatures and their connection to solar energetic particles}” from the International Space Sciences Institute (ISSI) Bern, Switzerland. \emph{Hinode} is a Japanese mission developed and launched by ISAS/JAXA, with NAOJ as domestic partner and NASA and UKSA as international partners. It is operated by these agencies in co-operation with ESA and NSC (Norway). CHIANTI is a collaborative project involving George Mason University, the University of Michigan (USA), University of Cambridge (UK) and NASA Goddard Space Flight Center (USA).

\section*{Data Availability}
For the data analysis, we used Interactive Data Language (IDL) and SolarSoftWare (SSW). \emph{Hinode}/EIS data is available at the \emph{Hinode} Science Data Centre Europe http://sdc.uio.no/sdc/. \emph{SDO}/AIA data is available at http://jsoc.stanford.edu/.



\end{document}